\documentclass[9pt,natbib]{sigplanconf}

\usepackage[utf8]{inputenc}
\usepackage[T1]{fontenc}
\usepackage{url}
\usepackage{color}
\usepackage{tabularx}
\usepackage{marvosym}
\usepackage{xspace}
\usepackage[nolist]{acronym}
\usepackage[normalem]{ulem}
\usepackage{algorithm}
\usepackage{algorithmicx}
\usepackage[noend]{algpseudocode}

\usepackage{listings}
\usepackage{mathptmx}
\usepackage{amsmath}
\usepackage{amssymb}
\usepackage{amsfonts}
\usepackage{amsthm}
\usepackage{graphicx}
\usepackage{qtree}
\usepackage{breakcites}
\usepackage{flushend}

\newcounter{foo}

\definecolor{orange}{RGB}{205,102,0}
\definecolor{darkgreen}{RGB}{0,200,0}
\definecolor{deepskyblue}{RGB}{0,154,205}

\newcommand{\ie}{i.\,e.\xspace}
\newcommand{\eg}{e.\,g.\xspace}

\providecommand*{\figrefname}{Figure~}
\providecommand*{\lstrefname}{Listing~}
\providecommand*{\tabrefname}{Table~}
\providecommand*{\secrefname}{Section~}
\providecommand*{\chaprefname}{Chapter~}
\providecommand*{\algrefname}{Algorithm~}
\providecommand*{\eqrefname}{Equation~}
\providecommand*{\linerefname}{Line~}

\theoremstyle{definition}

\begin{document}

\conferenceinfo{XLDI 2012}{September 9, 2012, Copenhagen, Denmark.}
\copyrightyear{2012}
\preprintfooter{SA paper for XLDI 2012 Workshop}
\titlebanner{DRAFT---Do not distribute}

\title{Enabling Operator Reordering in Data Flow Programs \\ Through 
Static Code Analysis}

\authorinfo{Fabian Hueske}{Technische Universit\"{a}t Berlin,
  Germany}{fabian.hueske@tu-berlin.de}

\authorinfo{Aljoscha Krettek}{Technische Universit\"{a}t Berlin,
  Germany}{aljoscha.krettek@campus.tu-berlin.de}

\authorinfo{Kostas Tzoumas}{Technische Universit\"{a}t Berlin,
  Germany}{kostas.tzoumas@tu-berlin.de}

\maketitle

\begin{abstract}
In many massively parallel data management platforms, programs are
represented as small imperative pieces of code connected in a data
flow. This popular abstraction makes it hard to apply algebraic
reordering techniques employed by relational DBMSs and other systems
that use an algebraic programming abstraction.
We present a code analysis technique based on reverse data and control 
flow analysis that discovers a set of properties from user code, 
which can be used to emulate algebraic optimizations in this setting.
\end{abstract}

\section{Introduction}
\label{sec:intro}

Motivated by the recent ``Big Data'' trend, a new breed of massively
parallel data processing systems has emerged. Examples of these
systems include MapReduce~\cite{dean.2004.osdi} and its open-source
implementation Hadoop~\cite{hadoop.2011.website},
Dryad~\cite{isard.2007.eurosys}, Hyracks~\cite{borkar.2011.icde}, and
our own Stratosphere system~\cite{battre.2010.socc}. These systems
typically expose to the programmer a {\em data flow} programming
model. Programs are composed as directed acyclic graphs (DAGs) of
operators, some of the latter typically being written in a
general-purpose imperative programming language. This model restricts
control flow only within the limits of operators, and permits only
dataflow-based communication between operators. Since operators can
only communicate with each other by passing sets of records in a
pre-defined hardwired manner, set-oriented execution and data
parallelism can be achieved.


Contrary to these systems, relational DBMSs, the traditional
workhorses for managing data at scale, are able to {\em optimize}
queries because they adopt an {\em algebraic} programming model based
on relational algebra. For example, a query optimizer is able to
transform the expression $\sigma_{R.X<3}(R \Join (S \Join T))$ to the
expression $(\sigma_{R.X<3}(R) \Join S) \Join T$, exploiting the
associativity and commutativity properties of selections and joins.

While algebraic reordering can lead to orders of magnitude faster
execution, it is not fully supported by modern parallel processing
systems, due to their non-algebraic programming models. Operators are
typically written in a general-purpose imperative language, and their
semantics are therefore hidden from the system. In our previous
work~\cite{hueske.2012.pvldb}, we bridged this gap by showing that
exposure of a handful of operator properties to the system can enable
reorderings that can simulate most algebraic reorderings used by
modern query optimizers. We discovered these properties using a
custom, shallow code analysis pass over the operators' code. Here we
describe this code analysis in detail, which we believe is of interest
by itself as an non-traditional use case of code analysis techniques.
We note that our techniques are applicable in the context of many data 
processing systems which support MapReduce-style UDFs such as parallel
programming models~\cite{dean.2004.osdi, battre.2010.socc}, 
higher-level languages~\cite{pig.web, chaiken.2008.pvldb}, and database 
systems~\cite{friedman.2009.vldb, greenplum.mr.web}.

\textbf{Related work:} In our previous work~\cite{hueske.2012.pvldb}
we describe and formally prove the conditions to reorder user-defined
operators. That paper also contains a more complete treatment of
related work. Here, we focus on more directly related
research. Manimal~\cite{jahani.2011.vldb} uses static code analysis of
MapReduce programs for the purpose of recommending possible
indexes. 
Our code analysis can be seen as an example of peephole
optimization~\cite{dragonbook}, and some of the concepts may bear
 similarity to techniques for loop optimization. However, we are not 
aware of code analysis being used before for the purpose of swapping 
imperative blocks of code to improve performance of data-intensive 
programs.

The rest of this paper is organized as
follows. Section~\ref{sec:udfReordering} describes the programming
model of our system, and introduces the reordering
technology. Section~\ref{sec:staticCodeAnalysis} discusses our code
analysis algorithm in detail. Finally, Section~\ref{sec:conclusion}
concludes and offers research directions.

\section{Data Flow Operator Reordering} \label{sec:udfReordering}



In our PACT programming model~\cite{battre.2010.socc}, a program $P$
is a DAG of sources, sinks, and operators which are connected by data
channels.  A source generates {\em records} and passes them to 
connected operators. A sink receives records from operators and 
serializes them into an output format. Records consist of fields of 
arbitrary types. To define an operator $O$, the programmer must
specify (i) a second-order function (SOF) signature, picked from a
pre-defined set of system second-order functions (currently {\em Map},
{\em Reduce}, \emph{Match}, \emph{Cross}, and \emph{CoGroup}), and
(ii) a first-order function (called user-defined function, UDF) that
is used as the parameter of the SOF. The model is strictly
second-order, in that a UDF is not allowed to call SOFs. The
intuition of this model is that the SOF defines a logical mapping
of the operator's input records into groups, and the UDF is invoked
once for each group. These UDF invocations are independent, and can be
thus scheduled on different nodes of a computing cluster.





\begin{figure}[thb]
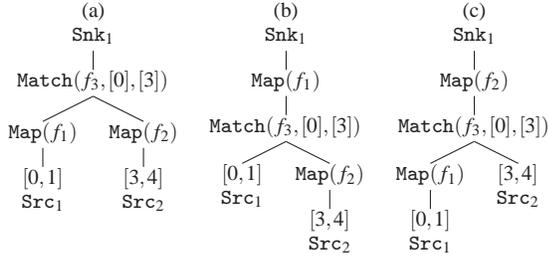

  \vspace{-4mm}
  \centering
  \small
  \begin{equation*}
  \Tree [.{(a)\\$\texttt{Snk}_1$} [.$\texttt{Match}(f_3,[0],[3])$ [.$\texttt{Map}(f_1)$ $[0,1]$\\$\texttt{Src}_1$ ] [.$\texttt{Map}(f_2)$ $[3,4]$\\$\texttt{Src}_2$ ] ] ]
~~~~~~ 
  \Tree [.{(b)\\$\texttt{Snk}_1$} [.$\texttt{Map}(f_1)$ [.$\texttt{Match}(f_3,[0],[3])$ $[0,1]$\\$\texttt{Src}_1$ [.$\texttt{Map}(f_2)$ $[3,4]$\\$\texttt{Src}_2$ ] ] ] ]
~~~~~~ 
  \Tree [.{(c)\\$\texttt{Snk}_1$} [.$\texttt{Map}(f_2)$ [.$\texttt{Match}(f_3,[0],[3])$ [.$\texttt{Map}(f_1)$ $[0,1]$\\$\texttt{Src}_1$ ] $[3,4]$\\$\texttt{Src}_2$ ] ] ]
  \end{equation*}
  \normalsize
  \vspace{-2mm}
  \caption{Example Data Flows: (a) original order, (b) first reordered alternative, (c) second reordered alternative}
   \vspace{-2mm}
  \label{fig:example}
\end{figure}

Figure~\ref{fig:example}(a) shows an example PACT program. The data
flow starts with two data sources $\texttt{Src}_1$ and $\texttt{Src}_2$ that provide records
which have the fields $[0,1]$ and $[3,4]$ set respectively (the
numbering is arbitrary). $\texttt{Src}_1$ feeds its data into a Map operator with
a UDF $f_1$. The Map SOF creates an independent group for each input
record, and $f_1$ is itself written in Java. UDF $f_1$ reads both
fields of its input record ($0$ and $1$), appends the sum of both
fields as field $2$, and emits the record. Similarly, the records of
$\texttt{Src}_2$ are forwarded to a Map operator with UDF $f_2$ which sums the
fields $3$ and $4$, appends the sum as field $5$ and emits the
record. The outputs of both Map operators are forwarded as inputs to a
Match operator with a UDF $f_3$ and the key field $[0]$ for the first
and $[3]$ for the second input.  The Match SOF creates a group for
each pair of records from both inputs that match on their key fields. $f_3$
merges the fields of both input records and emits the result. We give
the pseudo-code of all three user functions in the form of 3-address
code~\cite{dragonbook} below.
\vspace{0.2cm}\\
  \begin{tabular}{ll}
  \begin{minipage}{1.4in}
\begin{verbatim}
10: f1(InRec $ir)
11: $a:=getField($ir,0)
12: $b:=getField($ir,1)
13: $c:=$a + $b
14: $or:=copy($ir)
15: setField($or,2,$c)
16: emit($or)
\end{verbatim} 
\end{minipage} &
    \begin{minipage}{1.5in}
\begin{verbatim}
20: f2(InRec $ir)
21: $x:=getField($ir,3)
22: $y:=getField($ir,4)
23: $z:=$x + $y
24: $or:=create()
25: setField($or,3,$x)
26: setField($or,4,$y)
27: setField($or,5,$z)
28: emit($or)
\end{verbatim}
\end{minipage} \vspace{0.2cm}\\\hline
    \begin{minipage}{1.5in}
\vspace{0.2cm}
\begin{verbatim}
30: f3(InRec $ir1, InRec $ir2)
31: $or:=copy($ir1)
32: union($or,$ir2)
33: emit($or)
\end{verbatim} 
    \end{minipage} 
\vspace{2mm}
  \end{tabular} 

The pseudo-code shows the UDF API to process PACT records. The
user-functions \texttt{f1}, \texttt{f2}, and \texttt{f3} receive as
input one or two {\em input records} of type \texttt{InRec}. The only
way that a user function can emit an {\em output record} of type 
\texttt{OutRec} is by calling the \texttt{emit(OutRec)}
function. Output records can be either initialized as empty
(\texttt{OutRec create()}), or by copying an input record
(\texttt{OutRec copy(InRec)}). Records can be combined via the
function \texttt{void}
\texttt{union(}\texttt{OutRec,}\texttt{InRec)}. Fields can be read to
a variable via \texttt{Object getField}\texttt{(InRec,int)} addressed
by their position in the input record. The value of a field can be set
via \texttt{void setField(OutRec,int,Object)}. 
Note that our record API is based on basic operations and similar 
to other systems' APIs such as Apache Pig \cite{pig.web}.

Figures \ref{fig:example} (b) and (c) show potential reorderings of
the original data flow (a) where either $\texttt{Map}(f_1)$ or
$\texttt{Map}(f_2)$ has been reordered with
$\texttt{Match}(f_3,[0],[3])$. While data flow (b) is a valid
reordering, alternative (c) does not produce the same result as
(a). In previous work, we presented conditions for valid reorderings
of data flow operators centered around conflicts of operators on
fields~\cite{hueske.2012.pvldb}. For example, since we know that $f_1$
reads fields 0 and 1, and writes field 2, while $f_3$ reads fields 0
and 3, we can conclude that $f_1$ and $f_3$ only have a {\em read
  conflict} on field 0, and can thus be safely reordered.  UDFs that
have write conflicts cannot be reordered. This would be true if $f_1$
did not append the sum as field 2, but overwrote field 0 with the
sum. Additional complications arise from the way output records are
formed. Although on the first sight, $f_1$ and $f_2$ perform a very
similar operation, \ie, summing two fields and appending the result,
there is a fundamental difference. While $f_1$ creates its output
record by copying the input record (line \texttt{14}), $f_2$ creates
an empty output record (line \texttt{24}) and explicitly copies the
fields of the input record (lines \texttt{25,26}). The side effect of
creating an empty output record is that all fields of an input record
are implicitly removed from the output. By reordering
$\texttt{Map}(f_2)$ with $\texttt{Match}(f_3,[0],[3])$, the fields
$0$, $1$, and $2$ will get lost since $\texttt{Map}(f_2)$ does not
explicitly copy them into the newly created output record.

The information that needs to be extracted from the
user code in order to reason about reordering of operators is as follows.
The {\em read set} $R_f$ of a UDF $f$ is the set of fields from its input
data sets that {\em might} influence the UDF's output, \ie, fields that
are read and evaluated by $f$. The {\em write
  set} $W_f$ is the set of fields of the output data set that have
different values from the corresponding input field. The {\em emit
  cardinality bounds} $\lfloor EC_f \rfloor$ and $\lceil EC_f \rceil$
are lower and upper bounds for the number of records emitted per
invocation of $f$.
Reference~\cite{hueske.2012.pvldb} defines these properties
more formally, and provides conditions for reordering operators with
various SOFs given knowledge of these properties. In addition to change 
the order of operators, the optimizer can leverage these properties to 
avoid expensive data processing operations, \eg, a previously partitioned data 
set is still partitioned after a UDF was applied, if the partitioning 
fields were not modified by the UDF. Moreover, field projections 
can be pushed down based on read set information.

While it is very difficult to statically derive the exact properties by 
UDF code analysis in the general case, it is possible to conservatively 
approximate them. In reference~\cite{hueske.2012.pvldb} we discussed 
this static code analysis pass for the simple case of unary operators. 
In the next section, we provide the full algorithm that deals with the
additional complexity due to binary operators, and provide detailed
pseudo-code.

\section{Code Analysis Algorithm}
\label{sec:staticCodeAnalysis}

Our algorithm relies on a static code analysis (SCA) framework to get
the bytecode of the analyzed UDF, for example as typed three-address
code~\cite{dragonbook}. The framework must provide a control flow
graph (CFG) abstraction, in which each code statement is represented
by one node along with a function \textsc{Preds}($s$) that returns the
statements in the CFG that are ``true'' predecessors of statement $s$,
i.e., they are not both predecessors and descendants. Finally, the
framework must provide two methods $\textsc{Def-Use}(s, \texttt{\$v})$
and $\textsc{Use-Def}(s, \texttt{\$v})$ that represent the
Definition-Use chain of the variable \texttt{\$v} at statement $s$,
and the Use-Definition chain of variable \texttt{\$v} at statement $s$
respectively. Any SCA framework that provides these abstraction can be
used. 

\begin{algorithm}[t]
\small
\caption{Code analysis algorithm}
\label{alg:sca-generic}
\begin{algorithmic}[1]
\Function{Compute-Write-Set}{$f,O_f,E_f,C_f,P_f$} \label{line:write-set-start}
\State $W_f = E_f \cup P_f$ 
  \For{$i \in \Call{Inputs}{f}$}
    \If{$i \notin O_f$}
      $W_f = W_f \cup (\Call{Input-Fields}{f,i} \setminus C_f)$
    \EndIf
  \EndFor 
  \State \Return $W_f$ \label{line:write-set-end}
\EndFunction

\Function{Visit-UDF}{$f$}
  \State $R_f = \emptyset$ \label{line:rstart}
  \State $G=$ all statements of the form \texttt{g:\$t=getField(\$ir,$n$)}
  \For{\texttt{g} in $G$}
    \If{\Call{Def-Use}{\texttt{g}, \texttt{\$t}}$\neq\emptyset$}
      $R_f = R_f \cup \{n\}$ \label{line:rend}
    \EndIf
  \EndFor
  \State $E=$ all statements of the form \texttt{e:emit(\$or)} \label{line:emitstmts}
  \State $(O_f, E_f, C_f, P_f) = $ \Call{Visit-Stmt}{\textsc{Any}($E$), \texttt{\$or}} \label{line:emit-stmt-it-start}
  \For{\texttt{e} in $E$}
    \State $(O_e, E_e, C_e, P_e) = $ \Call{Visit-Stmt}{\texttt{e}, \texttt{\$or}}
    \State $(O_f, E_f, C_f, P_f) = $ \Call{Merge}{$(O_f, E_f, C_f, P_f), (O_e, E_e, C_e, P_e)$}
  \EndFor \label{line:emit-stmt-it-end}
  \State \Return $(R_f, O_f, E_f, C_f, P_f)$
\EndFunction

\Function{Visit-Stmt}{\texttt{s}, \texttt{\$or}}
  \If{$\textsc{visited}(\texttt{s}, \texttt{\$or})$} \label{line:memo}
    \State \Return \textsc{Memo-Sets}(\texttt{s}, \texttt{\$or})
  \EndIf
  \State $\textsc{Visited}(\texttt{s}, \texttt{\$or}) = \mathbf{true}$

  \If{\texttt{s} of the form \texttt{\$or = create()}} \label{line:create}
    \Return $(\emptyset, \emptyset, \emptyset, \emptyset)$
  \EndIf

  \If{\texttt{s} of the form \texttt{\$or = copy(\$ir)}} \label{line:copy}
    \State \Return $(\textsc{Input-Id}(\texttt{\$ir}), \emptyset, \emptyset, \emptyset)$
  \EndIf

  \State $P_s = $ \Call{Preds}{\texttt{s}} \label{line:setstmtstart}
  \State $(O_s, E_s, C_s, P_s) = $ \Call{Visit-Stmt}{\textsc{Any}($P_s$), \texttt{\$or}}
  \For{\texttt{p} in $P_s$}
    \State $(O_p, E_p, C_p, P_p) = $ \Call{Visit-Stmt}{\texttt{p}, \texttt{\$or}}
    \State $(O_s, E_s, C_s, P_s) = $ \Call{Merge}{$(O_s, E_s, C_s, P_s)$, $(O_p, E_p, C_p, P_p)$}
  \EndFor

  \If{\texttt{s} of the form \texttt{union(\$or, \$ir)}} \label{line:union}
    \State \Return $(O_{s} \cup \textsc{Input-Id}(\texttt{\$ir}) , E_{s}, C_{s}, P_{s})$

  \EndIf

  \If{\texttt{s} of the form \texttt{setField(\$or, $n$, \texttt{\$t})}} \label{line:setfield}
    \State $T =$\Call{Use-Def}{\texttt{s}, \texttt{\$t}}
    \If{all $\texttt{t} \in T$ of the form \texttt{\$t=getField(\$ir,$n$)}}
      \State \Return $(O_s, E_s, C_s \cup \{n\}, P_s)$
    \Else
      \State \Return $(O_s, E_s \cup \{n\}, C_s, P_s)$
    \EndIf
  \EndIf

  \If{\texttt{s} of the form \texttt{setField(\$or, $n$, null)}} \label{line:setnull}
    \State \Return $(O_s, E_s, C_s, P_s \cup \{n\})$ \label{line:setstmtend}
  \EndIf

\EndFunction
\Function{Merge}{$(O_1,E_1,C_1,P_1)$, $(O_2,E_2,C_2,P_2)$} \label{line:mergestart}
  \State $C = (C_1 \cap C_2) \cup \{x | x \in C_1, \textsc{Input-Id}(x) \in O_2 \} $ \\ $ ~~~~~~~~~~~~~~~~~~~~~~~~~~~~~\cup \{x | x \in C_2, \textsc{Input-Id}(x) \in O_1 \} $
  \State \Return $(O_1 \cap O_2, E_1 \cup E_2, C, P_1 \cup P_2)$ \label{line:mergeend}
\EndFunction
\end{algorithmic}
\end{algorithm}


The algorithm visits each UDF in a topological order implied by the
program DAG starting from the data sources. For each UDF $f$, the
function \textsc{Visit-Udf} of Algorithm~\ref{alg:sca-generic} is
invoked. First, we compute the read set $R_f$ of the UDF (lines
\ref{line:rstart}-\ref{line:rend}). For each statement of the form \texttt{\$t := getField(\$ir, n)}
that results in a valid use of variable \texttt{\$t}
(\textsc{Def-Use}($g$, \texttt{\$t})$\neq\emptyset$) we add field $n$ to $R_f$



Approximating the write set $W_f$ is more involved. We compute four
sets of integers that we eventually use to compute an approximation of
$W_f$. The {\em origin set} $O_f$ of UDF $f$ is a set of input ids. An
integer $o \in O_f$ means that all fields of the $o$-th input record
of $f$ are copied verbatim to the output. The {\em explicit
  modification set} $E_f$ contains fields that are modified and then
included in the output. We generally assume that fields are uniquely
numbered within the program (as in Figure~\ref{fig:example}). The {\em
  copy set} $C_f$ contains fields that are copied verbatim from one
input record to the output. Finally, the {\em projection set} $P_f$
contains fields that are projected from the output, by explicitly
being set to \texttt{null}. The write set is computed from these sets
using the function \textsc{Compute-Write-Set} (lines~\ref{line:write-set-start}-\ref{line:write-set-end}). 
All fields in $E_f$ and $P_f$ are explicitly modified or set to \texttt{null} and therefore in $W_f$. For inputs that are not in the origin set $O_f$, we add all fields of that input which are not in $C_f$, \ie, not explicitly copied.
  

To derive the four sets, function \textsc{Visit-Udf} finds all statements of the form \texttt{e:
  emit(\$or)}, which include the output record \texttt{\$or} in the
output (line \ref{line:emitstmts}). It then calls for each statement 
\texttt{e} the recursive function \textsc{Visit-Stmt} that recurses 
from statement \texttt{e} backwards in the control flow graph (lines \ref{line:emit-stmt-it-start}-\ref{line:emit-stmt-it-end}). The 
function performs a combination of reverse 
data flow and control flow analysis but does not change the values computed 
for statements once they have been determined. The function \textsc{Any}
returns an arbitrary element of a set.

The useful work is done in lines
\ref{line:setstmtstart}-\ref{line:setstmtend} of the algorithm. First,
the algorithm finds all predecessor statements of the current
statement, and recursively calls \textsc{Visit-Stmt}. The sets are
merged using the \textsc{Merge} function (lines
\ref{line:mergestart}-\ref{line:mergeend}). \textsc{Merge} provides a
conservative approximation of these sets, by creating maximal $E,P$
sets, and minimal $O,C$ sets. This guarantees that the data conflicts
that will arise are a superset of the true conflicts in the
program. When a statement of the form \texttt{setField(\$or, $n$,
  \texttt{null})} is found (line \ref{line:setnull}), field $n$ of the
output record is explicitly projected, and is thus added to the
projection set $P$. When a statement of the form
\texttt{setField(\$or, $n$, \texttt{\$t})} is found (line
\ref{line:setfield}), the \textsc{Use-Def} chain of \texttt{\$t} is
checked. If the temporary variable \texttt{\$t} came directly from
field $n$ of the input, it is added to the copy set $C$, otherwise it
is added to the explicit write set $E$. When we encounter a statement
of the form \texttt{\$or = create()} (line \ref{line:create}), we have
reached the creation point of the output record, where it is
initialized to the empty record. The recursion then ends. Another base
case is reaching a statement \texttt{\$or = copy(\$ir)} (line
\ref{line:copy}) where the output record is created by copying all
fields of the input record \texttt{\$ir}. This adds the input id of
record \texttt{\$ir} to the origin set $O$.  A \texttt{union}
statement (line \ref{line:union}) results in an inclusion of the input id of
the input record \texttt{\$ir} in the origin set $O$, and a further recursion for the
output record \texttt{\$or}. The algorithm maintains a memo table
\textsc{Memo-Sets} to support early exit of the recursion in the
presence of loops (line \ref{line:memo}). The memo table is implicitly
updated at every \texttt{return} statement of \textsc{Visit-Stmt}.

Function \textsc{Visit-Stmt} always terminates in the presence of
loops in the UDF code, since it will eventually find the statement
that creates the output record, or visit a previously seen
statement. This is due to \textsc{Preds} always exiting a loop after
visiting its first statement. Thus, loop bodies are only visited once
by the algorithm. The complexity of the algorithm is $O(en)$, where
$n$ is the size of the UDF code, and $e$ the number of emit
statements. This assumes that the Use-Def and Def-Use chains have been
precomputed.
 

The lower and upper bound on the emit cardinality of the UDF can be 
derived by another pass over the UDF code.
We determine the bounds for each emit statement $e$ and combine those to
derive the bounds of the UDF. For the lower bound $\lfloor EC_f
\rfloor$, we check whether there is a statement before statement $e$
that jumps to a statement after $e$. If there is none, the emit statement will always be executed and we set $\lfloor EC_f \rfloor=1$. If such a statement exists, statement $e$
could potentially be skipped during execution, so we set $\lfloor EC_f
\rfloor=0$. For the upper bound $\lceil EC_f \rceil$, we determine
whether there is a statement after $e$ that can jump to a
statement before $e$. If yes, the statement could be
executed several times during the UDF's execution, so we set
$\lceil EC_f \rceil = +\infty$. If such a statement does not exist,
statement $e$ can be executed at most once so we set $\lceil EC_f
\rceil = 1$. To combine the bounds we choose for the lower bound of
the UDF the highest lower bound over all emit statements and for the
upper bound the highest upper bound over all emit statements.

Our previous work~\cite{hueske.2012.pvldb} compares read and 
write sets which are automatically derived by our static
code analysis technique and from manually attached annotations. We show that our 
technique yields very precise estimations with only little loss of 
optimization potential. However, we note that the estimation quality depends 
on the programming style.

\section{Conclusions and Future Work}
\label{sec:conclusion}

We presented a shallow code analysis technique that operates on
data flow programs composed of imperative building blocks
(``operators''). The analysis is a hybrid of reverse data flow and
control flow analysis, and determines sets of record fields that express
the data conflicts of operators. These sets can be used to ``emulate''
algebraic reorderings in the dataflow program. Our techniques guarantee safety 
through conservatism and are applicable to many data processing systems that support 
UDFs. 
Future work includes research on intrusive user-code optimizations, 
\ie, modifying the code of UDFs, and on the effects that the use of functional 
programming languages to specify UDFs has on our approach and possible optimizations.

\acks This research was funded by the German Research Foundation under
grant FOR 1036. We would like to thank Volker Markl, our coauthors from previous work \cite{hueske.2012.pvldb}, and the members of the Stratosphere team.


\bibliographystyle{abbrv}
\small
\bibliography{paper}

\end{document}